\documentclass[%
amsmath,amssymb,
aps, reprint
]{revtex4-2}

\usepackage{hyperref} 
\usepackage{graphicx}
\usepackage{dcolumn}
\usepackage{bm}
\usepackage{graphicx}
\usepackage{epstopdf}
\usepackage{mathrsfs}
\usepackage{amssymb}
\usepackage{slashed}
\usepackage{verbatim}
\usepackage{color}
\usepackage{multirow}
\usepackage{amsmath}
\usepackage{epsfig}
\usepackage{ulem}
\usepackage{graphics}
\usepackage{indentfirst}
\usepackage{slashed}

\newcommand{\bea}{\begin{eqnarray}}
\newcommand{\eea}{\end{eqnarray}}
\newcommand{\beq}{\begin{equation}}
\newcommand{\eeq}{\end{equation}}

\newcommand{\vb}{\bar V}

\def\<{\langle}
\def\>{\rangle}

\def\nn{\nonumber}
\def\bZ {\mathbb{Z}}

\def\bZ {\mathbb{Z}}
\def\bM {\mathbb{M}}

\def\cO {{\mathcal O}}
\def\cT {{\mathcal T}}
\def\cR {{\mathcal R}}

\def\va{\vec{\alpha}}
\def\vb{\vec{\beta}}
\begin{document}

\preprint{APS/123-QED}

\title{Conformality and self-duality of $N_f=2$ QED$_3$} 

\author{Zhijin Li}
 \affiliation{Department of Physics, Yale University, New Haven, CT 06511} 

\date{\today} 

\begin{abstract}
We study the IR phase of three dimensional quantum electrodynamics (QED$_3$) coupled to $N_f=2$ flavors of two-component Dirac fermions,  which has been controversial for decades. This theory has been proposed to be self-dual  with  symmetry enhancement $(SU(2)_f\times U(1)_t )/\mathbb{Z}_2\rightarrow O(4)$ at the IR fixed point. We focus on the four-point correlator of monopole operators with unit topological charge of  $U(1)_t$.
We illustrate the $O(4)\rightarrow SU(2)_f\times U(1)_t $  branching rules based on an $O(4)$ symmetric positive structure in the monopole four-point crossing equations.
We use conformal bootstrap method to derive nonperturbative constraints on the CFT data and test the conformality and self-duality of $N_f=2$ QED$_3$. In particular we find the CFT data obtained from previous lattice simulations can be ruled out by introducing irrelevant assumptions in the spectrum, indicating the IR phase of $N_f=2$ QED$_3$ is not conformal. 
\end{abstract}
 
\maketitle


\section{Introduction}
Three dimensional $U(1)$ gauge theory coupled to $N_f$ two-component Dirac fermions has fundamental applications in layered condensed matter systems, including high temperature superconductors, topological insulators, etc.  The low energy limit of QED$_3$ shows  intriguing phenomena reminiscent to the 4D quantum chromodynamics: depending on the flavor number $N_f$, it can realize gauge confinement ($N_f=0$  \cite{Polyakov:1975rs, Polyakov:1976fu}), chiral symmetry breaking ($N_f\in (0, N_f^*)$) and conformal phase ($N_f\geq N_f^*$ \cite{Pisarski, Appelquist:1986fd, Appelquist:1988sr, Nash:1989xx}). The critical flavor number $N_f^*$ plays a key role in understanding the IR phases of QED$_3$ and its applications.

Various methods have been developed to estimate $N_f^*$. Perturbative approaches generically lead to a small $N_f^*$ in the range $0<N_f^* <10$ but it is hard to obtain a reliable estimation due to the strong coupling effect \cite{Appelquist:1988sr, Nash:1989xx, Maris:1996zg, Aitchison:1997ua, Appelquist:1999hr, Kubota:2001kk,   Franz_2003, Fischer:2004nq, Kotikov:2016wrb,  Kaveh_2005, Giombi:2015haa, DiPietro:2015taa, Giombi:2016fct, Zerf:2018csr, Herbut:2016ide, Gusynin:2016som, Christofi:2007ye, Janssen:2012pq, Braun:2014wja, Gukov:2016tnp}.  In lattice studies \cite{Hands:2002qt, Hands:2002dv, Hands:2004bh, Strouthos:2008kc, Thirmd}, the fermion bilinear condensation has been observed with $N_f=2$ which disappears with higher $N_f\geq 4$, indicating $2<N_f^*<4$. In contrast, lattice results from \cite{Karthik:2015sgq, Karthik:2016ppr, Qin:2017cqw, Karthik:2019mrr, Karthik:2020shl} suggest the $N_f=2$ QED$_3$ remains conformal in the low energy limit. The main contradiction among these results is whether the $N_f=2$ QED$_3$ has an IR fixed point.

The $N_f=2$ QED$_3$ also plays an important role in the 3D duality web \cite{Seiberg:2016gmd}. This theory has been proposed  to be self-dual in the IR with an enhanced $O(4)$ symmetry  \cite{Xu_2015}. Such self-duality is also obtained in \cite{Karch:2016sxi, Hsin:2016blu, Benini:2017dus, Wang:2017txt} from 3D fermion-fermion particle-vortex duality \cite{Son:2015xqa}.  
Moreover, based on the 3D fermion-boson duality, the $N_f=2$ QED$_3$ is further suggested to be dual to the easy-plane NCCP$^1$ model, which has been used to describe  the deconfined quantum critical point (DQCP) \cite{2004Sci, Senthil_2004x}. A crucial assumption in this scenario is that the $N_f=2$ QED$_3$ flows to an IR fixed point, i.e. $N_f^*<2$.  Besides, there could be two possibilities in the duality web \cite{Wang:2017txt, Senthil_2019}: the
{\it strong version} conjectures that the $O(4)$ symmetric fixed point is stable to all the perturbations (except a fermion mass term) allowed by the UV symmetries and all the theories in the duality web flow to the same IR fixed points; while the
{\it weak version} conjectures that the $O(4)$ symmetric fixed point is unstable to certain perturbations and only the $SU(2)$ symmetric QED$_3$ can flow to the IR fixed point.

Conformal bootstrap \cite{Rattazzi:2008pe, Poland:2018epd} provides a powerful nonperturbative approach for strongly coupled critical theories. It has been applied  to study $O(4)/SO(5)$ symmetric DQCPs \cite{Nakayama:2016jhq, Iliesiu:2018, unpb,Li:2018lyb} and   conformal QED$_3$ \cite{Chester:2016wrc, Chester:2017vdh}. In \cite{Li:2018lyb}  the author observed a new family of kinks in the $SO(N)$ vector bootstrap bounds for $N\geqslant6$, reminiscent to the well-known kinks related to the critical $O(N)$ vector models \cite{Kos:2013tga}, while the $SO(5)$ symmetric DQCP is just slightly below the conformal window. The kinks show interesting connections with conformal QED$_3$ and indicate a critical flavor number of QED$_3$: $2<N_f^*<4$. The bootstrap results provide evidence for the merger and annihilation of fixed points near $N_f^*$ \cite{Kubota:2001kk, Kaveh_2005, Gies:2005as, Kaplan:2009kr, Gorbenko:2018ncu}, though the underlying physics is entangled with the bootstrap bound coincidences among different symmetries \cite{Poland:2011ey, Li:2020bnb,Li:2020tsm}.
 
In this work we use conformal bootstrap to study the $N_f=2$ QED$_3$, which is the focal point of contradictions among a series of lattice studies in \cite{Hands:2002qt, Hands:2002dv, Hands:2004bh, Strouthos:2008kc, Thirmd} and \cite{Karthik:2015sgq, Karthik:2016ppr, Qin:2017cqw, Karthik:2019mrr, Karthik:2020shl}. The bootstrap bounds can provide strict and non-perturbative restrictions for the CFT data measured in the lattice simulations \cite{Karthik:2015sgq, Karthik:2016ppr, Qin:2017cqw, Karthik:2019mrr, Karthik:2020shl}. The results are in favor of a non-conformal IR phase of $N_f=2$ QED$_3$.
Together with the bootstrap results for the $N_f=4$ QED$_3$ \cite{Albayrak:2021} which support that the $N_f=4$ QED$_3$ is conformal and can be isolated into a closed region, the bootstrap results can provide a compelling estimation for the critical flavor number of QED$_3$.

\section{Monopole four-point correlator}
The $N_f=2$ QED$_3$ has a global symmetry $(SU(2)_f\times U(1)_t)/\bZ_2$, in which $SU(2)_f$ is the flavor symmetry while $U(1)_t$ is associated with the topological conserved current $j_\mu^t=\epsilon_{\mu\nu\rho} F^{\nu\rho}$. Operators charged under the  $U(1)_t$ symmetry are tied to configurations of the gauge field with nontrivial topologies,  namely the monopole operators \cite{Borokhov:2002ib}. The ``bare" charge 1 monopole $\mathcal{M}$ is not gauge invariant and it needs to be attached with a zero mode of the Dirac fermion $\Psi^i$ to construct gauge invariant monopoles $M_1\equiv \bar{\Psi}_i\mathcal{M}$. In consequence the monopoles $M_1$ also  form a spin-$\frac{1}{2}$ representation of $SU(2)_f$. They have been studied using perturbative approach \cite{Pufu:2013vpa, Dyer:2013fja, Chester:2017vdh} and the bootstrap method \cite{Chester:2016wrc, Chester:2017vdh}.

The self-duality of $N_f=2$ QED$_3$ is derived from 3D particle-vortex duality \cite{Xu_2015, Karch:2016sxi, Hsin:2016blu, Wang:2017txt}, which maps the fundamental fields of one theory to vortices of the other. For  $N_f=2$ QED$_3$,
the particle-vortex duality switches the $U(1)_t$ and the Cartan subgroup $U(1)_f\subset SU(2)_f$. The UV symmetry of the dual theory is $(SU(2)_t\times U(1)_f)/\bZ_2$. At the IR fixed point the global symmetry is a cover of the symmetries on both sides of the duality $(SU(2)_f\times SU(2)_t)/\bZ_2\cong SO(4)$. Moreover, the self-duality introduces a $\bZ_2$ symmetry exchanging $SU(2)_f$ and $SU(2)_t$, thus the true IR symmetry is $SO(4)\rtimes \bZ_2\cong O(4)$. The monopoles $M_1$ which form a spinor representation of $SU(2)_f$  transform as $(\frac{1}{2},\frac{1}{2})$ under $SU(2)_f\times SU(2)_t$.  
 
In above $O(4)$ symmetry enhancement, the four real components of monopoles $M_1$ which construct an $SU(2)$ fundamental representations are reset to form an $O(4)$ vector. Curiously the same pattern of ``symmetry enhancement" has been found  in the conformal four-point crossing equations  \cite{Li:2020bnb, Li:2020tsm} to explain the coincidences of bootstrap bounds with different global symmetries \cite{Poland:2011ey}. It should be reminded that this $O(4)$ symmetric positive structure in the monopole four-point correlator is purely algebraic and does not contain any dynamical information on the $O(4)$ DQCP. It can be used to clarify the $O(4)\rightarrow SU(2)_f\times U(1)_t$ branching rules in the QED$_3$ spectrum, which will be needed in our bootstrap studies.

Let us consider the four-point correlator of monopoles $M_1$.
It can be expanded into the $SU(2)_f\times U(1)_t$ irreducible representations \cite{Chester:2016wrc}
\bea
\langle M_1(x_1) M_1(x_2) M_1(x_3) M_1(x_4) \rangle = && \label{M4pt} \\
 \frac{1}{x_{12}^{2\Delta_M}x_{34}^{2\Delta_M}} \sum\limits_{r\in\{S,T,A\}}  && \sum\limits_{n\in\{0,1\}}
\cT_{n,r} G_{n,r}(u,v),   \nn
\eea 
in which the global symmetry indices have been  assumed implicitly. $\cT_{n,r}$ denotes the invariant four-point tensor associated with spin-$n$ representation of $SU(2)_f$ and  representation $r$ of $U(1)\simeq SO(2)_t$ including singlet ($S$), traceless symmetric ($T$) and anti-symmetric ($A$) \footnote{The $S$ and $A$ representations of $SO(2)$ are isomorphic, however, they have opposite spin selection rules in the $M_1\times M_1$ OPE \cite{Chester:2016wrc}. }.  Variables $u,v$ are the conformal invariant cross ratios.

\begin{table}[b]
\caption{\label{tab1}%
$SU(2)_f\times SO(2)_t$ and $O(4)$ representations in the monopole four-point correlator. Operators in the $SO(2)_t$ $T$ sectors carry charge 2 of the topological $U(1)_t$ symmetry and do not have definite parity charges, in the sense that operators in this sector may have both even or odd parity charges for different components. }
\begin{ruledtabular}
\begin{tabular}{cccccc}
$SU(2)_f$ & $SO(2)_t$ &
Spin & Parity & DOF &
$O(4)$   \\
\hline
0& $S$ & odd & $-$ & $\mathbf{1}$ & $(1,0)+(0,1)$   \\
1& $S$ & even & $-$ &$\mathbf{3}$  & $(1,1)$  \\
0& $A$ & even & $+$ & $\mathbf{1}$ & $(0,0)$  \\
1& $A$ & odd & $+$ & $\mathbf{3}$ & $(1,0)+(0,1)$  \\
0& $T$ & odd & $\cdot$ & $\mathbf{2}$ & $(1,0)+(0,1)$  \\
1& $T$ & even & $\cdot$ & $\mathbf{6}$ & $(1,1)$  
\end{tabular}
\end{ruledtabular}
\end{table}

The correlator (\ref{M4pt}) satisfies the $SU(2)_f\times SO(2)_t$ symmetric crossing equations \cite{Chester:2016wrc}
\beq
\bM^{Q}_{6\times 6} \cdot \mathbf{1}_{6\times 1}=\mathbf{0}_{6\times1}, \label{creq1}
\eeq
and the  square matrix $\bM^Q_{6\times 6}$ is
{ 
\bea
\left(
\begin{array}{cccccc}
 0 & 0 & F^-_{0,A} & F^-_{1,A} & F^-_{0,T} & F^-_{1,T}  \\
 -F^-_{0,S} & -F^-_{1,S} & -F^-_{0,A} & -F^-_{1,A} & 0 & 0 \\
 -F^-_{0,S} & 3 F^-_{1,S} & F^-_{0,A} & -3 F^-_{1,A} & -2 F^-_{0,T} & 6 F^-_{1,T} \\
 0 & 0 & F^+_{0,A} & -3 F^+_{1,A} & F^+_{0,T} & -3 F^+_{1,T} \\
 -F^+_{0,S} & 3 F^+_{1,S} & -F^+_{0,A} & 3 F^+_{1,A} & 0 & 0 \\
 -F^+_{0,S} & -F^+_{1,S} & F^+_{0,A} & F^+_{1,A} & -2 F^+_{0,T} & -2 F^+_{1,T}
\end{array}
\right),\nn
\eea
}
with correlation functions $F^\pm_{n,r}$
\beq
F^\pm_{n,r}\equiv v^{\Delta_{M_1}} G_{n,r}(u,v)\mp u^{\Delta_{M_1}} G_{n,r}(v,u).
\eeq
The crossing equations of $O(4)$ vector are \cite{Rattazzi:2010yc}
\beq
\bM^{O}_{3\times 3} \cdot \mathbf{1}_{3\times 1}=\mathbf{0}_{3\times1}, \label{creq2}
\eeq
where 
\bea
\bM^{O}_{3\times 3} = \left(
\begin{array}{ccc}
 0 & F^-_T & -F^-_A \\
 F^-_S & \frac{1}{2}F^-_T & F^-_A \\
 F^+_S & -\frac{3}{2}F^+_T & -F^+_A \\
\end{array}
\right). \label{MO}
\eea
Though the crossing equations (\ref{creq1}) and (\ref{creq2}) are rather different, 
there is a transformation $\mathscr{T}_{3\times 6}$ 
\beq
\mathscr{T}_{3\times 6}=\left(
\begin{array}{cccccc}
 -\frac{1}{3} & \frac{2}{3} & 0 & -1 & -\frac{2}{3} & \frac{4}{3} \\
 \frac{1}{3} & \frac{1}{3} & 1 & 1 & \frac{2}{3} & \frac{2}{3} \\
 -\frac{1}{3} & -1 & 1 & -1 & -\frac{2}{3} & -2 \\
\end{array}
\right),
\eeq
 which surprisingly connects the two crossing equations 
\bea 
\mathscr{T}_{3\times 6}\cdot  \bM^{Q}_{6\times 6} &=& \label{TrsM} \\ 
&& \hspace{-2cm}\left(
\begin{array}{cccccc}
 -\frac{1}{3}F^-_{0,S} & \frac{2 }{3} F^-_{1,S}& 0 & -F^-_{1,A} & -\frac{2}{3} F^-_{0,T} & \frac{4}{3}F^-_{1,T} \\
 \frac{1}{3}F^-_{0,S} & \frac{1}{3}F^-_{1,S} & F^-_{0,A} & F^-_{1,A} & \frac{2}{3} F^-_{0,T} & \frac{2 }{3} F^-_{1,T}\\
 -\frac{1}{3}F^+_{0,S} & -F^+_{1,S} & F^+_{0,A} & -F^+_{1,A} & -\frac{2}{3} F^+_{0,T} & -2 F^+_{1,T} \\
\end{array}
\right). \nn
\eea
Above matrix is essentially the $\bM^{O}_{3\times 3}$ in (\ref{MO}), with replicated columns multiplied by positive rescaling factors. The many-to-one maps from $SU(2)_f\times SO(2)_t$ to $O(4)$ representations  are provided 
in the Table \ref{tab1}. 
Here we would like to add a few comments about the parity charges in different $SO(2)_t$ sectors. Operators in the $S/A$ sectors are neutral under the topological $SO(2)_t$ symmetry. Their parity charges depend on the flavor number $N_f$. 
As discussed in \cite{Chester:2016wrc}, the monopole operator $M_1$ in $N_f=2$ QED$_3$ is pseudo-real with a reality condition
\beq
(M_1^{ia})^\dagger\propto\epsilon_{ij}\epsilon^{ab}M_1^{ib}, \label{rlt}
\eeq
where $i,j$ ($a,b$) denote the $SU(2)_f$ ($SO(2)_t$) indices. The  monopole operator $M_1$ transforms under the spacetime reflection operation $\cR$:
$\cR M_1(x) \cR=M_1^\dagger(-x)$.
From the reality condition (\ref{rlt}) and reflection positivity restriction  $\langle M_1(x)M_1(-x)^\dagger\rangle>0$, one can fix the normalization of the monopole operator $M_1$ 
\beq
\langle M_1^{ia}(x_1)M_1^{jb}(x_2)\rangle=\frac{\epsilon^{ij}\epsilon^{ab}}{x_{12}^{2\Delta_{M_1}}}.
\eeq 
According to the above normalization, the $(0,A)$ sector with $SO(2)_t$ tensor structure $\epsilon^{ab}$ has even parity, while the $(0,S)$ sector has opposite parity charge. For instances, the unit operator and stress tensor operator appear in the $(0,A)$ sector which are parity even, while the topological $SO(2)_t$ conserved current $j_\mu^t$ with odd parity charge appears in the sector $(0, S)$.
In the QED$_3$ with even $N_f/2$, e.g. $N_f=4$, the monopole operators $M_1'$ are real instead of pseudo-real, and they satisfy different reality conditions which lead to opposite parity charges of operators in the $S/A$ sectors.
\footnote{The parity charges of the $SO(2)_t$ $S/A$ sectors in Table \ref{tab1} are opposite to those in \cite{Chester:2016wrc}. In \cite{Chester:2016wrc} operators in the $A/S$ sector are considered to be parity odd/even in QED$_3$  with general flavors $N_f$; however, this does not work for pseudo-real monopole operators, e.g. the $M_1$ in $N_f=2$ QED$_3$ studied in this work. We thank Shai Chester and Silviu Pufu for discussions on this issue.} In the $SO(2)_t$ $T$ sectors, operators $\cO_q$ have  charge $|q|=2$ of the topological $SO(2)_t$ symmetry. The parity symmetry transforms the monopole operators $\cO_q$ to $\cO_{-q}$. These monopole operators do not have definite parity charges, instead different components of the monopoles $\cO_q$ transform oppositely under parity symmetry.

Table \ref{tab1} shows how the $SU(2)_f\times SO(2)_t$ representations are combined into $O(4)$ representations. 
This suggests the $SU(2)_f\times U(1)_t$ crossing equations actually have an $O(4)$  symmetric positive structure, which can lead to the $O(4)$ symmetry enhancement in the bootstrap bounds and $SO(N)$-ization of four-point correlators \cite{Li:2020bnb, Li:2020tsm}. 
The subtlety in this symmetry enhancement is that the monopoles carrying topological $U(1)_t\subset SU(2)_t$ charges can be combined with $U(1)_t$ neutral operators to form $SU(2)_t$ representations. Specifically, the singlet $(0,A)$ remains singlet of $O(4)$ \footnote{ It is important for bootstrap studies that the $O(4)$ singlets only comes from the singlets of $SU(2)_f\times SO(2)_t$, otherwise the lowest $O(4)$ singlet could be from an $SU(2)_f\times SO(2)_t$ non-singlet and breaks the irrelevant condition in this channel. For instance, in the fermion bilinear $(1,S)$ bootstrap, we cannot simply introduce an irrelevant condition in the $SU(2)_f$ singlet sector as there could be relevant operators which are the singlet of $SU(2)_f$ while non-singlet of  $O(4)$.}; the $O(4)$ conserved current consists of the conserved currents associated with $SU(2)_f$ and $U(1)_t$ symmetries, and a charge 2 monopole in the $(0, T)$ sector. The fermion bilinears in the $(1, S)$ sector construct an $O(4)$ traceless symmetric scalar after combining with a $(1, T)$ monopole, which agrees with the results in \cite{Benini:2017dus}.


\section{Bootstrap bounds}
\begin{figure}[b]
\includegraphics[width=1\linewidth]{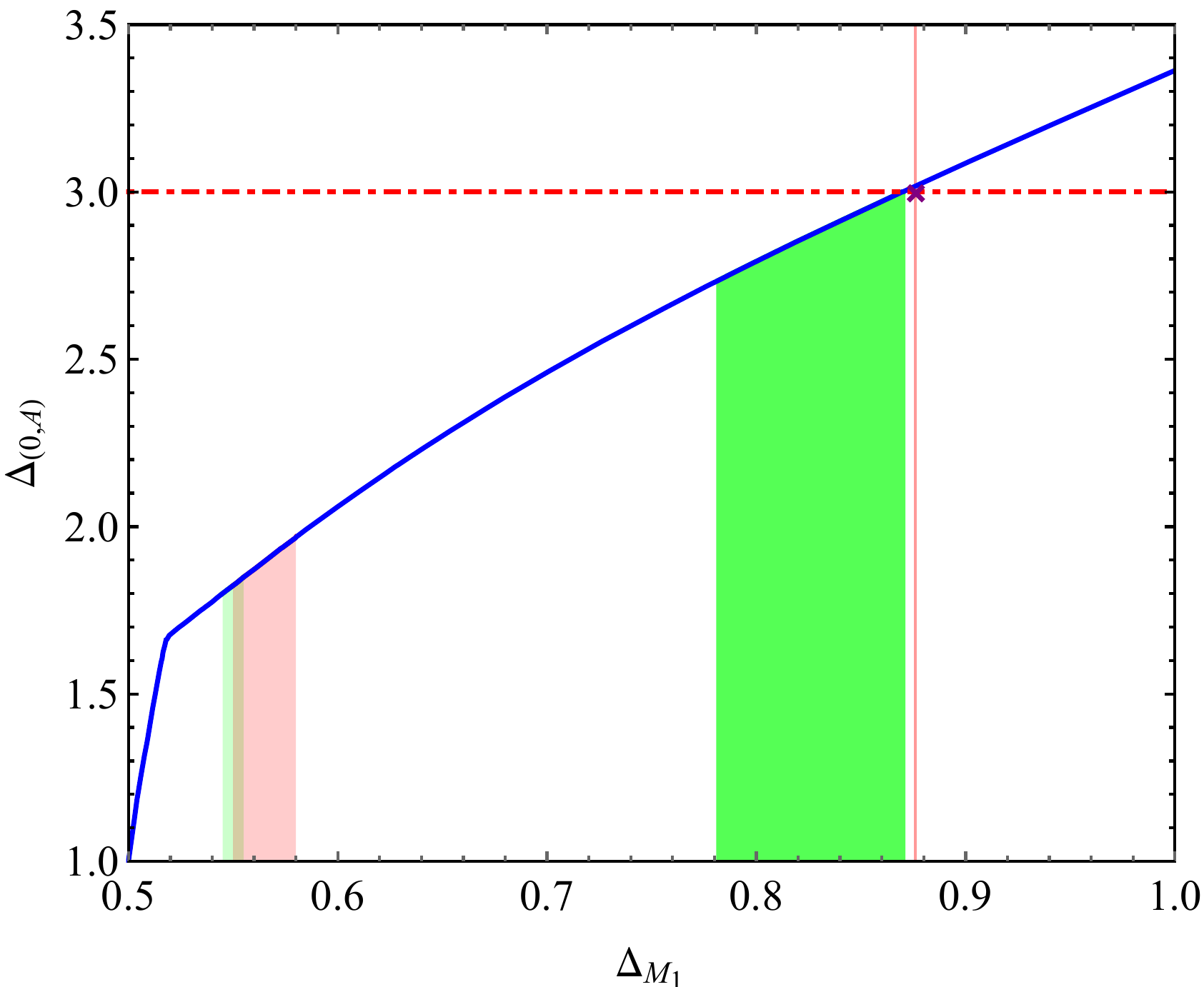} 
\caption{\label{fig1} Blue line: upper bound on the scaling dimension of the lowest singlet scalar $\Delta_{(0,A)}$ in the monopole bootstrap with $\Lambda=31$, which coincides with the singlet bound in the $O(4)$ vector bootstrap. Light green (red) region: lattice result  of the BH (EPJQ) model  $\Delta_{V}=0.550(5)$ ($\Delta_{V}=0.565(15)$)  \cite{Qin:2017cqw}. Dark green region:  direct QED$_3$ simulation $\Delta_{M_1}=0.826(44)$ \cite{Karthik:2019mrr}. Vertical red line: minimum $\Delta_{M_1}$ ($\Delta_{M_1}=0.876$) at $\Lambda=51$ with a gap assumption $\Delta_{(0,A)}\geq3$. }
\end{figure}

We bootstrap the monopole four-point correlator (\ref{M4pt}) to derive constraints on CFTs with an $SU(2)\times U(1)$ symmetry. 
In Fig. \ref{fig1} we show the upper bound on the scaling dimension of the lowest singlet scalar in the $(0, A)$ sector. The $SU(2)\times SO(2)$  singlet bound exactly coincides with the singlet bound from $O(4)$ vector bootstrap. The kink in the bound near $\Delta_{M_1}\sim 0.52$ gives the critical $O(4)$ vector model, whose bootstrap solution was firstly discovered in \cite{Kos:2013tga}. The coincidence of $SU(2)\times SO(2)$ and $O(4)$ bootstrap bounds can be explained by the hidden relation between their crossing equations given by (\ref{TrsM}). The transformation $\mathscr{T}_{3\times 6}$ actually maps the $SU(2)\times SO(2)$ bootstrap problems to the $O(4)$ vector bootstrap problems in a way consistent with positive conditions required by bootstrap algorithm. 
When bootstrapping non-singlet scalars, the $O(4)$ positive structure in $\bM^Q_{6\times 6}$ can be broken by the gap conditions in the bootstrap setup, which lead to different bootstrap results, e.g., the bootstrap bounds on $\Delta_T$ and $\Delta_{(1,S)}$ in Fig. \ref{fig1b}. 
See Appendix \ref{BDCS} for more details.  
 
\begin{figure}[b]
\includegraphics[width=1\linewidth]{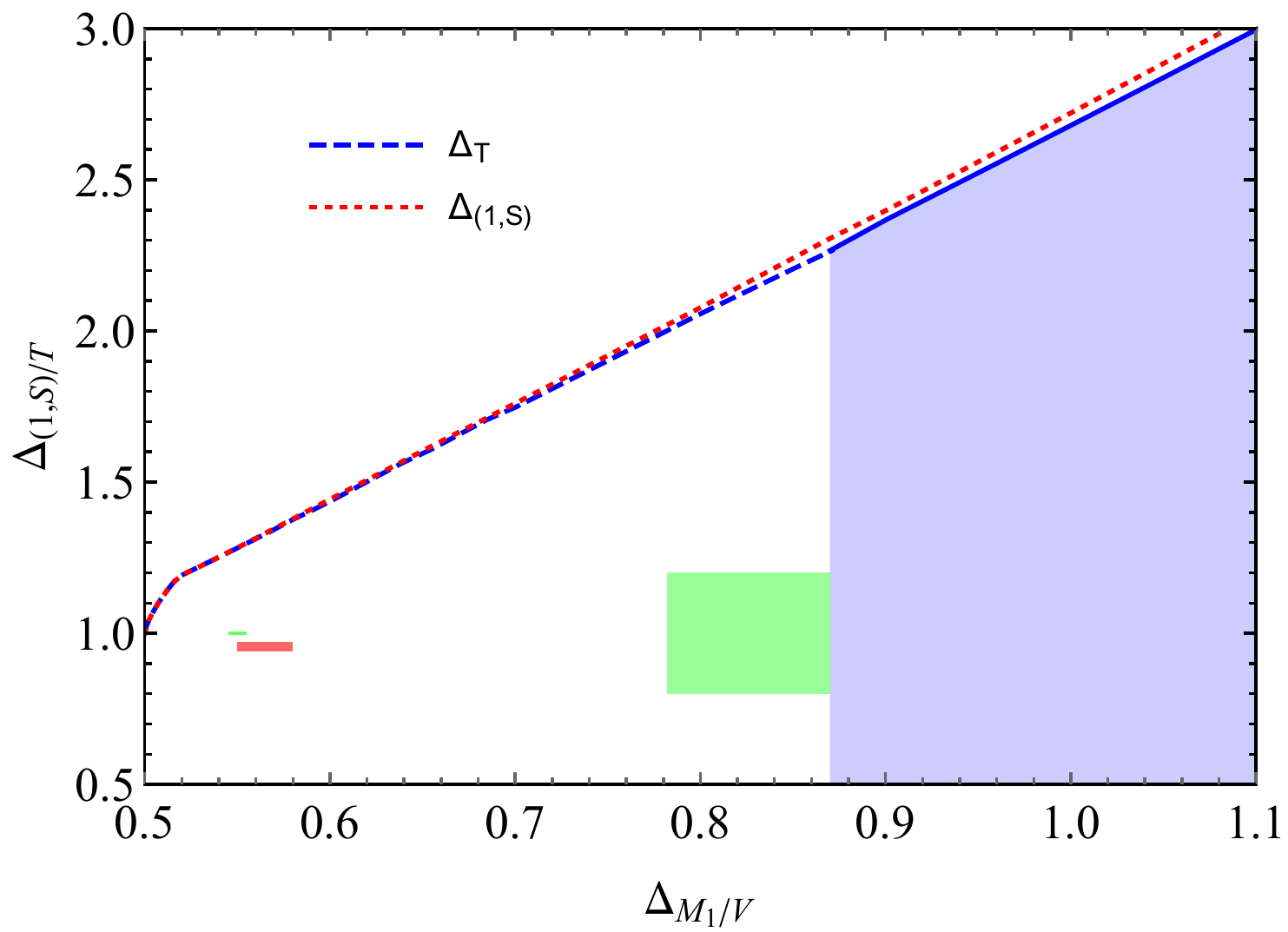} 
\caption{\label{fig1b} Dashed lines: upper bounds on $\Delta_T$ and $\Delta_{(1,S)}$. Blue shadowed region: bootstrap allowed region of $(\Delta_V,\Delta_T)$ with a gap assumption $\Delta_S>3$. The red and two green shadow regions denote estimations from lattice simulations.}
\end{figure}

In $N_f=2$ QED$_3$, the lowest singlet scalar in the $(0, A)$ sector is a mixing of the four-fermion operator $(\bar\Psi_i \Psi^i)^2$ and the gauge kinetic term $F_{\mu\nu}F^{\mu\nu}$ \footnote{It is critical that the $(0,A)$ sector is parity even. There is a parity odd fermion bilinear singlet scalar which is relevant. }. This operator has dimension $4$ in the UV and can receive notable correction in the IR. It has been assumed to be irrelevant in the IR to realize a fixed point of $N_f=2$ QED$_3$ and its dualities. QED$_3$ like CFTs with a relevant singlet scalar relate to the universality classes of QED$_3$-GNY model instead. In the ``merger and annihilation" scenario, the two theories merge at the critical flavor $N_f^*$, and the lowest singlet scalar approaches marginality condition $\Delta_S=3$. According to the conformal perturbation theory, the two fixed points move to complex plan and become non-unitary once  the lowest singlet scalar crosses marginality condition \footnote{In contrast, it is possible that certain non-singlet scalar crosses marginality condition at critical flavor $N_f^*$ while the fixed points remain on both sides of $N_f^*$ and interchange stability, like the cubic and $O(N)$ vector models. See \cite{Gukov:2016tnp, Gorbenko:2018ncu} for more comprehensive discussions.}. In lattice simulations, with a relevant singlet scalar it needs fine tuning to reach the IR fixed point. In particular, the DQCP requires the lowest  singlet scalar being irrelevant otherwise the fixed point with enhanced $O(4)$ symmetry cannot be realized \cite{Wang:2017txt}. According to the bootstrap bound in Fig. \ref{fig1}, this gap assumption leads to a constraint $\Delta_{M_1}\geq 0.870$ at $\Lambda=31$, and it increases to $\Delta_{M_1}\geq 0.876$  using higher numerical precision with $\Lambda=51$ \footnote{ The parameter $\Lambda$ is the maximum number of derivatives in the linear functional, which is the major factor determining the numerical precision of bootstrap results \cite{SD2015qma}.  $\Lambda=31$ is assumed in this work unless otherwise specified.}.
 
Lattice results on the IR phase of $N_f=2$ QED$_3$ remain controversial. In \cite{Hands:2002qt, Hands:2002dv, Hands:2004bh, Strouthos:2008kc} the chiral symmetry breaking has been observed which suggests this theory is not conformal. However, in \cite{Karthik:2015sgq, Karthik:2016ppr, Qin:2017cqw} the authors observed conformal behavior in the large distance limit and the scaling dimensions of some low-lying operators have been estimated. Specifically, in \cite{Qin:2017cqw} the  $N_f=2$ QED$_3$ has been simulated using bilayer honeycomb (BH) model. The results also support the $O(4)$ symmetry enhancement and the duality with the NCCP$^1$ model realized by the easy-plane-J-Q (EPJQ) model. The scaling dimensions of $O(4)$ vector and traceless symmetric scalars are $\Delta_{V}=0.550(5), ~\Delta_{T}=1.000(5)$ in the BH model and $\Delta_{V}=0.565(15), ~\Delta_{T}=0.955(15)$ in the EPJQ model. According to the bootstrap bound, a unitary CFT with $\Delta_{V}$ in this range needs a strongly relevant singlet scalar $\Delta_S<2.0$, which will necessarily break the RG flow to the IR fixed point with an enhanced $O(4)$ symmetry \cite{DPunpb}. 
In a lattice simulation of  $N_f=2$ QED$_3$ \cite{Karthik:2019mrr}, the scaling dimension of $M_1$ is estimated in the range $\Delta_{M_1}=0.826(44)$, and the upper bound is $\Delta_{M_1}\leq 0.87$, which is ``conspiratorially" near the lower bound of bootstrap result $\Delta_{M_1}\geq 0.87$ at $\Lambda=31$ and $\Delta_{M_1}\geq 0.876$ at $\Lambda=51$  by requiring an irrelevant lowest singlet scalar! Here conformal bootstrap provides quantitatively elaborate consistent check for a unitary CFT. On the other hand, more precise lattice simulations will definitely be  helpful in this comparison. Note the two lattice estimations on $\Delta_V$ \cite{Qin:2017cqw, Karthik:2019mrr} are significantly different, while the estimations on $\Delta_T$ from \cite{Qin:2017cqw} are consistent with the estimation in \cite{Karthik:2016ppr} on the fermion bilinears $\Delta_{(1,S)}=1.0(2)$.  
The bootstrap bound with condition $\Delta_{S}\geqslant3$ strongly (mildly) excludes the lattice results in \cite{Qin:2017cqw} (\cite{Karthik:2016ppr, Karthik:2019mrr}).


\begin{figure}[b]
\includegraphics[width=1\linewidth]{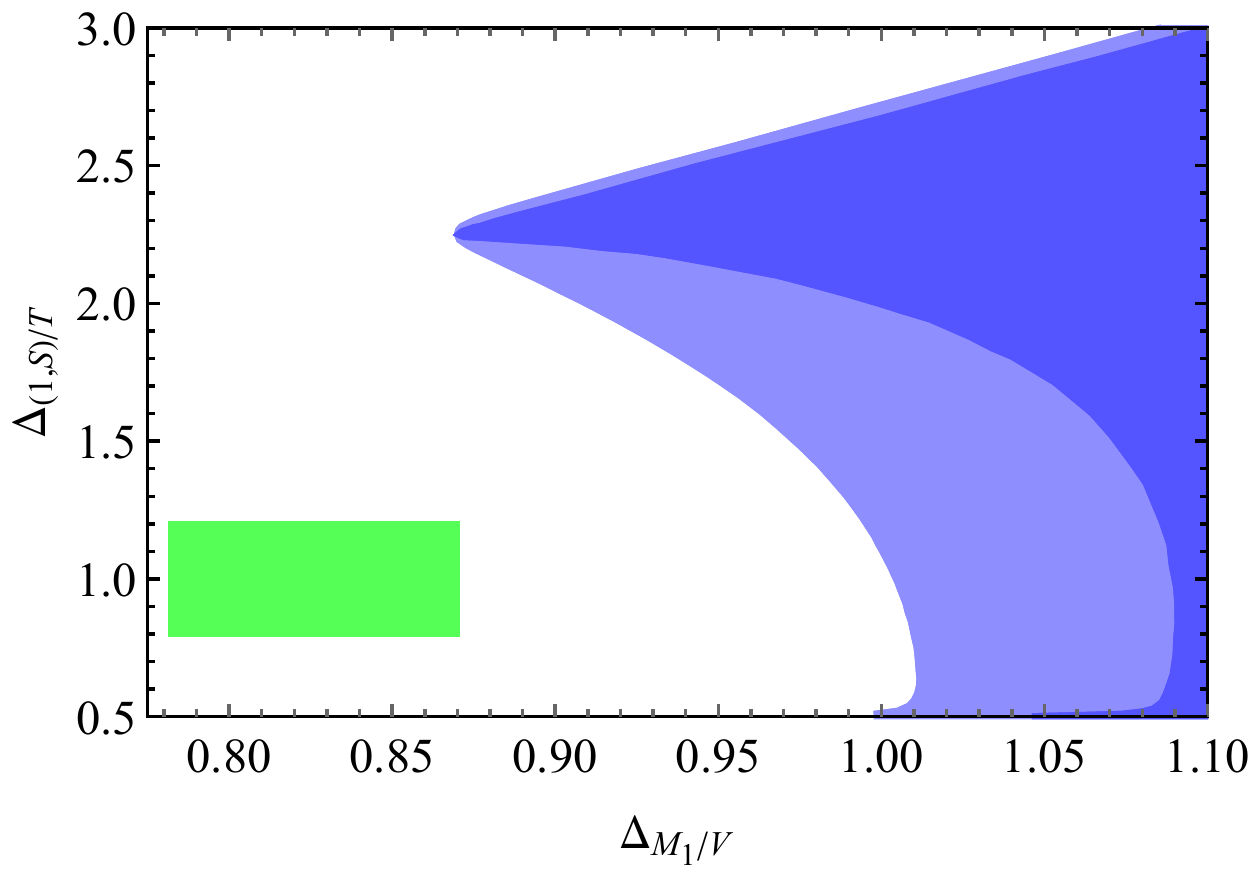} 
\caption{\label{fig2} Light blue: $\Delta_{M_1}\sim \Delta_{(1,S)}$ under assumptions $\Delta_{(0,A)}\geqslant3$ and the second $(1,S)$ scalar is irrelevant. Dark blue: $\Delta_V \sim \Delta_T$ from $O(4)$ vector bootstrap with assumptions $\Delta_S\geqslant3$ and the second traceless symmetric ($T$) scalar is irrelevant. Green region: lattice results \cite{Karthik:2016ppr, Karthik:2019mrr} on $\Delta_{M_1}\sim \Delta_{(1,S)}$.}
\end{figure}

The monopole bootstrap can also generate constraints on the scaling dimension of fermion bilinear mass term ${\bar\Psi}_1\Psi^1-{\bar\Psi}_2\Psi^2$ in the $(1, S)$ sector shown in Figs. \ref{fig1b} and \ref{fig2}. The results will be particularly useful to test the {\it strong version} of the duality web, which conjectures that both the $U(1)\times U(1)$ and $SU(2)$ symmetric QED$_3$ flow to the same IR fixed point. The $U(1)\times U(1)$ symmetric QED$_3$ allows couplings which break $SU(2)_f$ to $U(1)_f$, e.g. scalars in $(1,S)$ sector. The lowest scalar in $(1,S)$ sector is the fermion bilinear which is tuned to zero at the critical point, while the second lowest scalar in this sector needs to be irrelevant to realize IR fixed point in the $U(1)\times U(1)$ symmetric theory.
This operator has dimension 4 in the UV and is conjectured to be irrelevant in the IR.  
The $U(1)\times U(1)$ QED$_3$ 
is interested in the  duality web of DQCPs. Firstly the derivation of the $N_f=2$ QED$_3$ self-duality takes the two fermions separately, and the self-duality is supposed to work for the $U(1)\times U(1)$ symmetric theories \cite{Wang:2017txt}.
Moreover, the easy-plane NCCP$^{1}$ model has a $U(1)\times U(1)$ symmetry. According to the {\it strong version} of duality web, terms breaking $SU(2)_f$ to $U(1)_f$ should also be allowed in the fermionic theory. 
In terms of the $O(4)$ representations, this suggests the lowest scalar in the $(1,1)$ sector is the tuning parameter while the second lowest scalar is irrelevant. Similar symmetry-breaking terms also appear in the $(2,2)$ representation of $O(4)$. However, they only appear in the OPE of $(1,1)$ scalars instead of $O(4)$ vectors and their bootstrap study is provided in Appendix \ref{bt4}.

In Fig. \ref{fig2} we show the bootstrap allowed region  for $\Delta_{M_1} \sim \Delta_{(1,S)}$ (light blue) with assumptions required by the strong version of the dualities, that both the lowest singlet scalar in $(0, A)$ sector and the second lowest  scalar in $(1, S)$ sector are irrelevant. 
If we further impose $O(4)$ symmetry, more regions in the $\Delta_{V} \sim \Delta_{T}$ plan can be carved out (dark blue). The bootstrap bounds in Fig. \ref{fig2} can be applied to the lattice results \cite{Karthik:2016ppr, Qin:2017cqw}, in which the only tuning parameter is the fermion mass term and the IR phase is stable under other perturbations. This suggests the next operator in the $(1,S)$ or $T$ sector should be irrelevant in their lattice setup.
For the $N_f=2$ QED$_3$ with fermion mass operator $\Delta_{(1,S)}=1.0(2)$, the bootstrap bound requires $\Delta_{M_1}\geqslant 0.99$ without imposing the $O(4)$ symmetry, while the lower bound increases to $\Delta_V\geqslant 1.08$ after introducing $O(4)$ symmetry. In both scenarios, the lattice results \cite{Karthik:2016ppr, Qin:2017cqw, Karthik:2019mrr} locate in the regions  notably away from the bootstrap bounds and are clearly excluded. 

\section{Conclusions}
We have provided nonperturbative constraints on the IR phase of $N_f=2$ QED$_3$ using conformal bootstrap. The results show that the CFT data of $N_f=2$ QED$_3$ measured from lattice simulations  \cite{Karthik:2015sgq, Karthik:2016ppr, Qin:2017cqw, Karthik:2019mrr, Karthik:2020shl} are not consistent with bootstrap bounds both with and without an enhanced $O(4)$ symmetry, which suggest the phase transitions observed in these simulations are not truly continuous, resolving the contradiction on the IR phase of $N_f=2$ QED$_3$ among a series of lattice studies 
\cite{Hands:2002qt, Hands:2002dv, Hands:2004bh, Strouthos:2008kc, Thirmd} and \cite{Karthik:2015sgq, Karthik:2016ppr, Qin:2017cqw, Karthik:2019mrr, Karthik:2020shl}.
Our results support that the $N_f=2$ QED$_3$ is not conformal. On the other hand, the bootstrap study of $N_f=4$ QED$_3$ \cite{Albayrak:2021} provides strong evidence for its conformal IR phase, by showing that part of perturbative CFT data of this theory provides a consistent solution to the crossing equations and can be isolated into a closed region. The two results provide a compelling answer for the critical flavor number of QED$_3$: $2<N_f^*<4$,  consistent with  the lattice simulations \cite{Hands:2002qt, Hands:2002dv, Hands:2004bh, Strouthos:2008kc} and the observation in the bootstrap results \cite{Li:2018lyb}. 


\begin{acknowledgments}
The author would like to thank Soner Albayrak, Meng Cheng, Rajeev Erramilli, David Poland and Yuan Xin for stimulating discussions. 
The author is also grateful to Nikhil Karthik for helpful communications on the lattice studies. 
This work  is supported by Simons Foundation grant 488651 (Simons Collaboration on the Nonperturbative Bootstrap) and DOE grant no.\ DE-SC0017660. Computations in this work were carried out on the Yale Grace computing cluster, supported by the facilities and staff of the Yale University Faculty of Sciences High Performance Computing Center.
\end{acknowledgments}

\appendix

\section{$SU(2)\times SO(2)$ and $O(4)$ bootstrap bound coincidence} \label{BDCS}
In this section we apply the algebraic relation (\ref{TrsM}) to conformal bootstrap. The relation (\ref{TrsM}) suggests that the two crossing equations $\bM^{Q}_{6\times 6}$
and $\bM^{O}_{3\times 3}$ have the same positive structure, which leads to coincidences in certain bootstrap bounds. In particular, the upper bound on the scaling dimension of $SU(2)\times SO(2)$ singlet scalar in the $M_1\times M_1$ OPE is saturated by the $O(4)$ symmetric four-point correlators, with $O(4)\rightarrow SU(2)\times SO(2)$ branching rules consistent with the proposed $N_f=2$ QED$_3$ self-duality.

To bootstrap the four-point correlator of the monopoles $M_1$ , we start from its crossing equation (\ref{creq1}). The conformal correlation functions $F_{n,r}^\pm$ are expanded in terms of 3D conformal partial waves  $g_{\Delta,\ell}(u,v)$ \cite{Dolan:2000ut, Dolan:2011dv}:
\beq
G_{n,r}(u,v)=\sum\limits_{\cO\in (n,r)}\lambda_{M_1M_1\cO}^2 g_{\Delta_\cO,\ell_\cO}(u,v).
\eeq
Accordingly, the conformal partial wave expansion of the crossing equations (\ref{creq1}) is given by  $\bM^{Q}_{6\times 6}$ with a replacement
\beq
F_{n,r}^\pm\rightarrow F_{\Delta_\cO,\ell_\cO}^\pm
\eeq
where 
\beq
F^\pm_{\Delta_\cO,\ell_\cO}\equiv v^{\Delta_{M_1}} g_{\Delta_\cO,\ell_\cO}(u,v)\mp u^{\Delta_{M_1}} g_{\Delta_\cO,\ell_\cO}(v,u).
\eeq
Now let us consider the following bootstrap problem: what is the maximum scaling dimension $\Delta_S^*$ of the lowest non-unit singlet scalar that can appear in the $M_1\times M_1$ OPE and satisfy above crossing equations and unitarity conditions?
According to the standard bootstrap algorithm, this bootstrap problem is equivalent to find  linear functionals $\vb$ for a given $\Delta_{M_1}$, which satisfy positive conditions when applied on conformal blocks $F_{\Delta,\ell}^\pm(u,v)$
\begin{widetext}
\bea
\vb\cdot \bM^{Q}_{6\times 6}= \left(\beta _1~\beta _2 ~ \beta _3 ~ \beta _4 ~ \beta _5 ~ \beta _6\right) &\cdot &\left(
\begin{array}{cccccc}
 0 & 0 & F^-_{\Delta,\ell} & F^-_{\Delta,\ell} & F^-_{\Delta,\ell} & F^-_{\Delta,\ell}  \\
 -F^-_{\Delta,\ell} & -F^-_{\Delta,\ell} & -F^-_{\Delta,\ell} & -F^-_{\Delta,\ell} & 0 & 0 \\
 -F^-_{\Delta,\ell} & 3 F^-_{\Delta,\ell} & F^-_{\Delta,\ell} & -3 F^-_{\Delta,\ell} & -2 F^-_{\Delta,\ell} & 6 F^-_{\Delta,\ell} \\
 0 & 0 & F^+_{\Delta,\ell} & -3 F^+_{\Delta,\ell} & F^+_{\Delta,\ell} & -3 F^+_{\Delta,\ell} \\
 -F^+_{\Delta,\ell} & 3 F^+_{\Delta,\ell} & -F^+_{\Delta,\ell} & 3 F^+_{\Delta,\ell} & 0 & 0 \\
 -F^+_{\Delta,\ell} & -F^+_{\Delta,\ell} & F^+_{\Delta,\ell} & F^+_{\Delta,\ell} & -2 F^+_{\Delta,\ell} & -2 F^+_{\Delta,\ell} \\
\end{array}
\right)    \nn \\
&=&(\beta_{0,S} ~ \beta_{1,S} ~ \beta_{0,A} ~ \beta_{1,A} ~  \beta_{0,T} ~ \beta_{1,T})
\succcurlyeq 0,  ~~~  \forall \Delta \geqslant \Delta_S^* \text{ or unitary bound.}  \label{sumatrix}
\eea
\end{widetext}
The parameters $(\Delta_{M_1},\Delta_{S}^*)$ can be excluded if the actions $\beta_{n,r}$ of linear functionals $\vb$ are strictly positive. In the optimal case when the conditions $\beta_{n,r}=0$  are satisfied for a set of spectrum, one obtains extremal solutions to the crossing equations and the boundary of bootstrap bound is reached. 

The $O(4)$ vector bootstrap problem can be defined similarly. To obtain the maximum scaling dimension of the lowest non-unit  $O(4)$ singlet scalar appearing in the OPE of an $O(4)$ vector $\phi_i$, we aim to find the linear functionals $\va$ which satisfy
\begin{widetext}
\bea
\va\cdot \bM^{O}_{3\times 3}=\left(
 \alpha _1 ~ \alpha _2 ~ \alpha _3 
\right) \cdot  \left(
\begin{array}{ccc}
 0 & F^-_{\Delta,\ell} & -F^-_{\Delta,\ell} \\
 F^-_{\Delta,\ell} & \frac{1}{2}F^-_{\Delta,\ell} & F^-_{\Delta,\ell} \\
 F^+_{\Delta,\ell} & -\frac{3}{2}F^+_{\Delta,\ell} & -F^+_{\Delta,\ell} \\
\end{array}
\right) =  (\alpha_S ~ \alpha_T ~\alpha_A)    \succcurlyeq (0~0~0),  
~~~\forall \Delta  \geqslant \Delta_S^* \text{ or unitary bound.~~} \label{onmatrix}
\eea
\end{widetext}

Now the algebraic relation (\ref{TrsM}) between $SU(2)\times SO(2)$ and $O(4)$ crossing equations plays a critical role for above two seemingly unrelated bootstrap problems (\ref{sumatrix}) and (\ref{onmatrix}). Assuming we have already obtained the $O(4)$ bootstrap linear functionals $\vec{\alpha}^*$ satisfying the positive conditions in  (\ref{onmatrix}), we can construct linear functional $\vb^*$ for the $SU(2)\times SO(2)$ bootstrap
\bea
\vb^*_{1\times 6}\equiv \va^*_{1\times 3}\cdot \mathscr{T}_{3\times6}.
\eea
Its action on the crossing equation matrix $\bM^{Q}_{6\times 6}$ is
\begin{widetext}
\bea
\vb^*_{1\times6}\cdot \bM^{Q}_{6\times 6} =\va_{1\times 3}^*\cdot \mathscr{T}_{3\times6} \cdot \bM^{Q}_{6\times 6} 
=\left(\frac{1}{3}\alpha^*_A,\frac{2}{3}\alpha^*_T,\alpha^*_S, \alpha^*_A,\frac{2}{3}\alpha^*_A,\frac{4}{3}\alpha^*_T\right). \label{act1} 
\eea
\end{widetext}
This is exactly the $O(4)$ vector bootstrap action with $O(4)\rightarrow SU(2)\times SO(2)$ branching rules
\bea
S_{O(4)}&\rightarrow& (0, A), \\
T_{O(4)}&\rightarrow& (1, S) \oplus (1,T), \label{branchT} \\
A_{O(4)}&\rightarrow& (0, S)\oplus (1, A) \oplus (0, T),
\eea
associated with positive recombination coefficients $\left(\frac{1}{3},\frac{2}{3},1, 1,\frac{2}{3},\frac{4}{3}\right)$. 
Therefore it has exactly the same positive property as the $O(4)$ action $(\alpha^*_S, \alpha^*_T, \alpha^*_A)$. Combining with the bootstrap algorithm, this suggests that any data $(\Delta_{M_1},\Delta_S^*)$ that is excluded by the $O(4)$ vector bootstrap, is also excluded by the  $M_1$ monopole bootstrap. This results to the following relation on the bootstrap upper bound $\Delta_S^c$
\beq
\Delta_{S}^c|_{M_1} \leq  \Delta_{S}^c|_{O(4)}. \label{neq1}
\eeq
On the other hand, on the boundary of bootstrap we obtain extremal solutions to the crossing equations, and any $O(4)$ symmetric solution can be decomposed to the $SU(2)\times SO(2)$ symmetric solutions, therefore we also have
\beq
\Delta_{S}^c|_{M_1} \geq  \Delta_{S}^c|_{O(4)}. \label{neq2}
\eeq
Relations (\ref{neq1}) and (\ref{neq2}) leads to the identity
\beq
\Delta_{S}^c|_{M_1} = \Delta_{S}^c|_{O(4)},
\eeq
which proves the bootstrap bound coincidence in the singlet sector observed in Fig. \ref{fig1}. Moreover, Eq. (\ref{act1}) shows the extremal solution of the $SU(2)\times SO(2)$ singlet bootstrap bound is $O(4)$ symmetric!

In general there is no such coincidence in the non-singlet bootstrap bounds. Let us consider the upper bounds on the scalars in the $(1,1)$ representation of  $O(4)$ and $(1,S)$ representation of $SU(2)\times SO(2)$ for example. In the former case, we require the action of linear functional is non-negative for $\forall \Delta_{\cO\in (1,1)} \geqslant \Delta_T^*$  in the bootstrap equations (\ref{onmatrix}), while in the later case we require the action of linear functional is non-negative for $\forall \Delta_{\cO\in(1,S)} \geqslant \Delta_{(1,S)}^*$ and  $\forall \Delta_{\cO\in(1,T)} \geqslant 0.5 \text{ (unitary bound)}$  in the bootstrap equations (\ref{sumatrix}). 
Considering the branching relation (\ref{branchT}), above requirement in $O(4)$ bootstrap is stronger than that in $SU(2)\times SO(2)$ bootstrap, consequently the bootstrap bound on $O(4)$ $(1,1)$ scalar is stronger than the bound on $SU(2)\times SO(2)$ $(1, S)$ scalar unless we impose an extra constraint that the $(1,S)$ and $(1,T)$ have the same scaling dimension.

\section{Bootstrap bounds from $O(4)$ $(1,1)$ bootstrap} \label{bt4}
In the main body of this paper, we have shown the results of $N_f=2$ QED$_3$ monopole bootstrap. In the self-duality scenario, the charge 1 monopole operator $M_1$ forms a vector representation $(\frac{1}{2},\frac{1}{2})$ of $O(4)$ symmetry. The monopole bootstrap implementation has access to the $(0,0), (1,0)+(0,1)$ and $(1,1)$ sectors. The results suggest conformal bootstrap can provide nontrivial constraints for some fundamental questions on strongly coupled gauge theories. 
In this part we present the $O(4)$ $(1,1)$ bootstrap results, which provide less conclusive but still interesting constraints on the $O(4)$ DQCP.

$O(4)$ representations that can appear in the $(1,1)$ bootstrap include
\bea
(1,1)\otimes (1,1) &=&(0,0)\oplus (1,1)\oplus(2,2)\oplus
 (1,0)+(0,1)  \nn \\
&&\oplus (1,2)+(2,1)\oplus (2,0)+(0,2). \label{OPEt2} 
\eea
Crossing equations of the four-point correlator of the $(1,1)$ scalar can be found in \cite{Reehorst:2020phk}. 
The scalar operators in $(2,2)$ representation carry four $O(4)$ indices $T_{(ijkl)}$. Some of its components can appear in the easy-plane NCCP$^1$ model, e.g., $\sum_{i=1}^2\sum_{j=3}^4 T_{(iijj)}$ corresponding to the N\'eel-VBS anisotropy and  $\sum_{i=1}^2 T_{(iiii)}$ in the microscopic models on the square lattice \cite{Wang:2017txt}. However, these components are forbidden for the $SU(2)$ symmetric QED$_3$. If the $O(4)$ symmetric IR fixed point has a relevant scalar  $T_{(ijkl)}$, as proposed in the weak version of $O(4)$ duality web, then the easy-plane model is unstable under the perturbations mentioned above, while the $N_f=2$ QED$_3$ can flow to it. In the $SU(2)$ symmetric  QED$_3$, the perturbations in (\ref{OPEt2}) allowed by the symmetry are $O(4)$ singlet or the $(0,2)$ representation. If there are relevant operators in these sectors, then even the $SU(2)$ symmetric QED$_3$ cannot flow to the IR fixed point and the weak version of $O(4)$ DQCP cannot be realized in the lattice simulation. Generally it is hard to compute scaling dimensions of the operators in these complicated representations. Fortunately, conformal bootstrap can provide useful information on these problems.

\begin{figure}[b]
\includegraphics[scale=0.75]{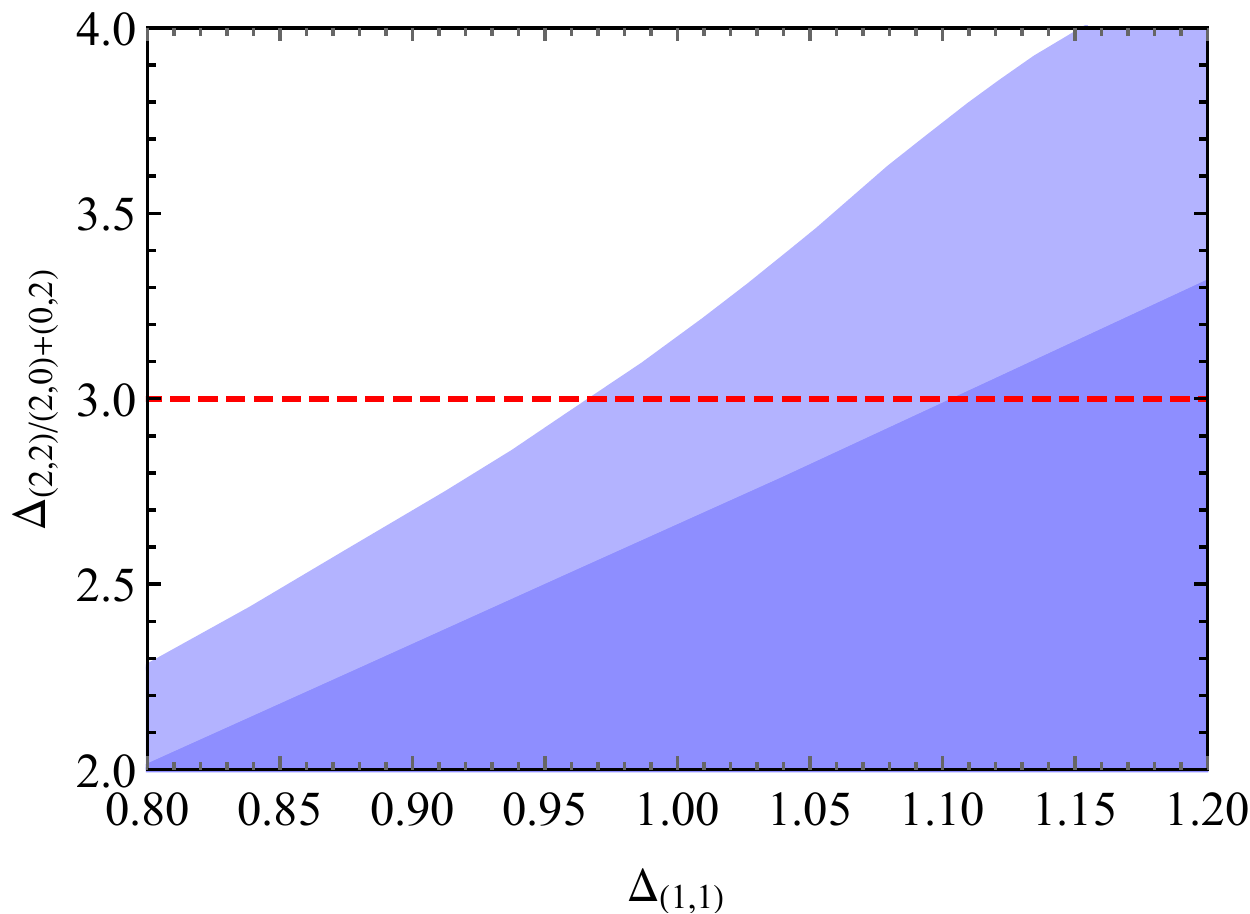} 
\caption{\label{fig3} Upper bounds on the scaling dimensions of the lowest scalars  in the $(2, 2)$ (darker blue) and $(2,0)+(0,2)$ (lighter blue) representations of $O(4)$. To get the bound of $(2,0)+(0,2)$, we have used extra assumptions $\Delta_{0,0}\geq 3$ and the lowest scalar in $(1,1)$ representation is given by the external scalar in the  four-point correlator.}
\end{figure}

In Fig. \ref{fig3} we show the bootstrap bounds on the lowest scalars in $(2,2)$ and $(2,0)+(0,2)$ representations. The bootstrap bounds suggest the lowest $(2,2)$ scalar is relevant in the range $\Delta_{(1,1)}<1.1$, while the lowest $(2,0)+(0,2)$ scalar is relevant with $\Delta_{(1,1)}<0.95$.
In the lattice simulation \cite{Qin:2017cqw}, scaling dimension of the $O(4)$ (1,1) scalar 
is estimated at $\Delta_{(1,1)}=1.000(5)$ in BH model and $\Delta_{(1,1)}=0.955(15)$ in the EPJQ model. The lowest $(2,2)$ scalar in the theory has to be relevant and the strong version of $O(4)$ duality is excluded, while the lowest $(0,2)$ scalar can be marginally irrelevant. The lattice results in \cite{Karthik:2016ppr}  has a more significant error bar on the scaling dimension of $O(4)$ $(1,1)$ scalar: $\Delta_{(1,1)}=1.0(2)$. Bootstrap results suggestion only in the range $\Delta_{(1,1)}>1.1$
the lowest $(2,2)$ can be irrelevant, which is necessary for the strong version of $O(4)$ duality web.


\bibliography{Nf2QED3}

\end{document}